\documentstyle[11pt,newpasp,twoside,epsf]{article}
\def\edcomment#1{\iffalse\marginpar{\raggedright\sl#1\/}\else\relax\fi}
\reversemarginpar
\def\ecs{erg~cm$^{-2}$s$^{-1}$}

\begin{document}
\title{Long and faint GRBs in archival BeppoSAX-WFC data}
 \author{J.J.M. in 't Zand, J. Heise}
 \affil{Space Research Organization Netherlands \&
 University Utrecht}
  \author{R.M. Kippen}
 \affil{Los Alamos National Laboratory}
 \author{P.M. Woods}
 \affil{NASA Marshall Space Flight Center}
 \author{C. Guidorzi}
 \affil{University of Ferrara}
 \author{E. Montanari, F. Frontera}
 \affil{CNR IASF Bologna \& University of Ferrara}

\begin{abstract}
In an ongoing effort to search for X-ray transients on diverse time
scales in the data archive of the BeppoSAX Wide Field Cameras we have
identified four GRBs that are long (between 540 and at least 2550 s)
and faint (with 3 events having peak fluxes below
$7\times10^{-9}$~\ecs\ in 2--10 keV).  Three of the events were
covered {\em and} detected by BATSE, suggesting that there may be a
substantial number of bursts still hidden in the BATSE database. A
spectral analysis of the three events shows that, although soft, they are
not typical X-ray flashes. The fourth event (not covered by BATSE) is
within 3\arcmin\ coincident with a double-lobed radio galaxy. The
chance probability is estimated at 10$^{-3}$. If real, the association
would be remarkable.
\end{abstract}

\section{Introduction}

The longest {\em continuously active} GRB detected with BATSE is
GRB~971208 (Connaughton et al. 1997; Giblin et al. 2002) which
exhibits a single fast-rise exponential-decay profile lasting a couple
of thousand seconds. Such a long duration is an interesting feature
because it touches upon questions as when the prompt emission ends and
the afterglow emission starts, and on the detect-ability of bursts at
extreme redshifts. In principle prolonged (and fainter) GRB emission
is easier to detect with a device that has (some) imaging capability
because there will not be confusion with variations in the background
radiation. Particularly this applies to coded aperture devices because
they employ many detector pixels per point source and, thus, are also not
susceptible to 'hot' pixels.

The BeppoSAX Wide Field (coded aperture) Cameras (WFCs; Jager et
al. 1997) are able to pick up transient events that are as short as a
few tens of milliseconds (from SGR~1900+14) and as long as months to
years (from X-ray binaries). The algorithms that are employed to
search for new transient sources differ per time scale, in an attempt
to simultaneously optimize processing speed and sensitivity. On time
scales between 1~s and 1~min, photon counts from the whole detector
are searched for increases in a manner typical for triggering
algorithms in non-imaging gamma-ray devices. The typical on-axis
sensitivity is $2\times10^{-9}$~\ecs\ (2--10 keV). On time scales
between one BeppoSAX orbit (1.7~hr) and one BeppoSAX pointing (mostly
about 1~d), full Iterative-Removal-of-Sources (IROS; Hammersley et
al. 1992) reconstructions are performed and the resulting images
searched for unidentified point sources. The typical sensivity is
$(1-5)\times10^{-10}$~\ecs. Yet another imaging algorithm is applied
for time scales between 1~min and 1~hr. We here report on four
transients which were detected in the IROS reconstructions of orbital
data stretches. Three of these were covered by BATSE. Thus, we are
able to measure the broad-band spectrum of these extreme examples of
GRBs.

\begin{figure}
\plottwo{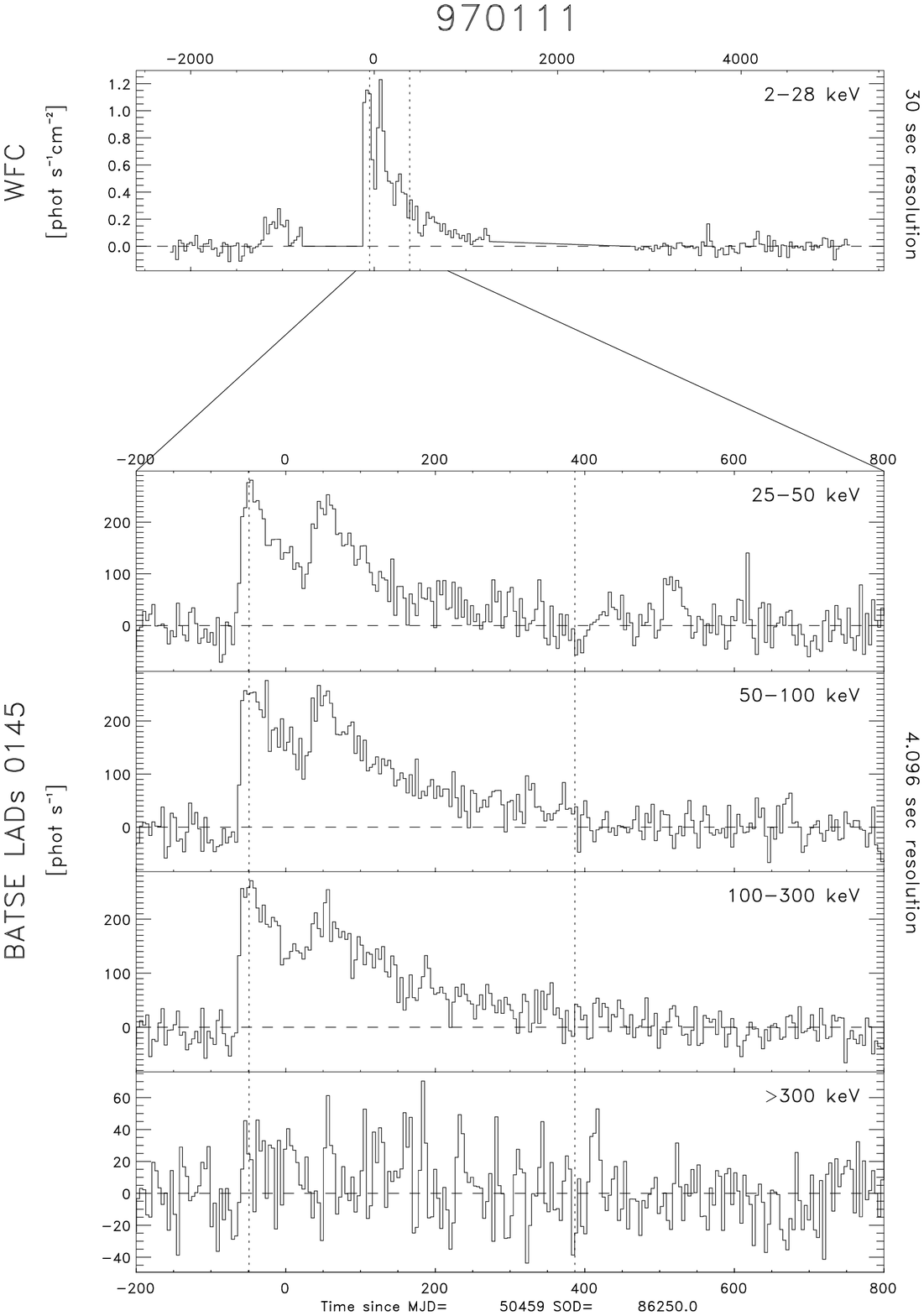}{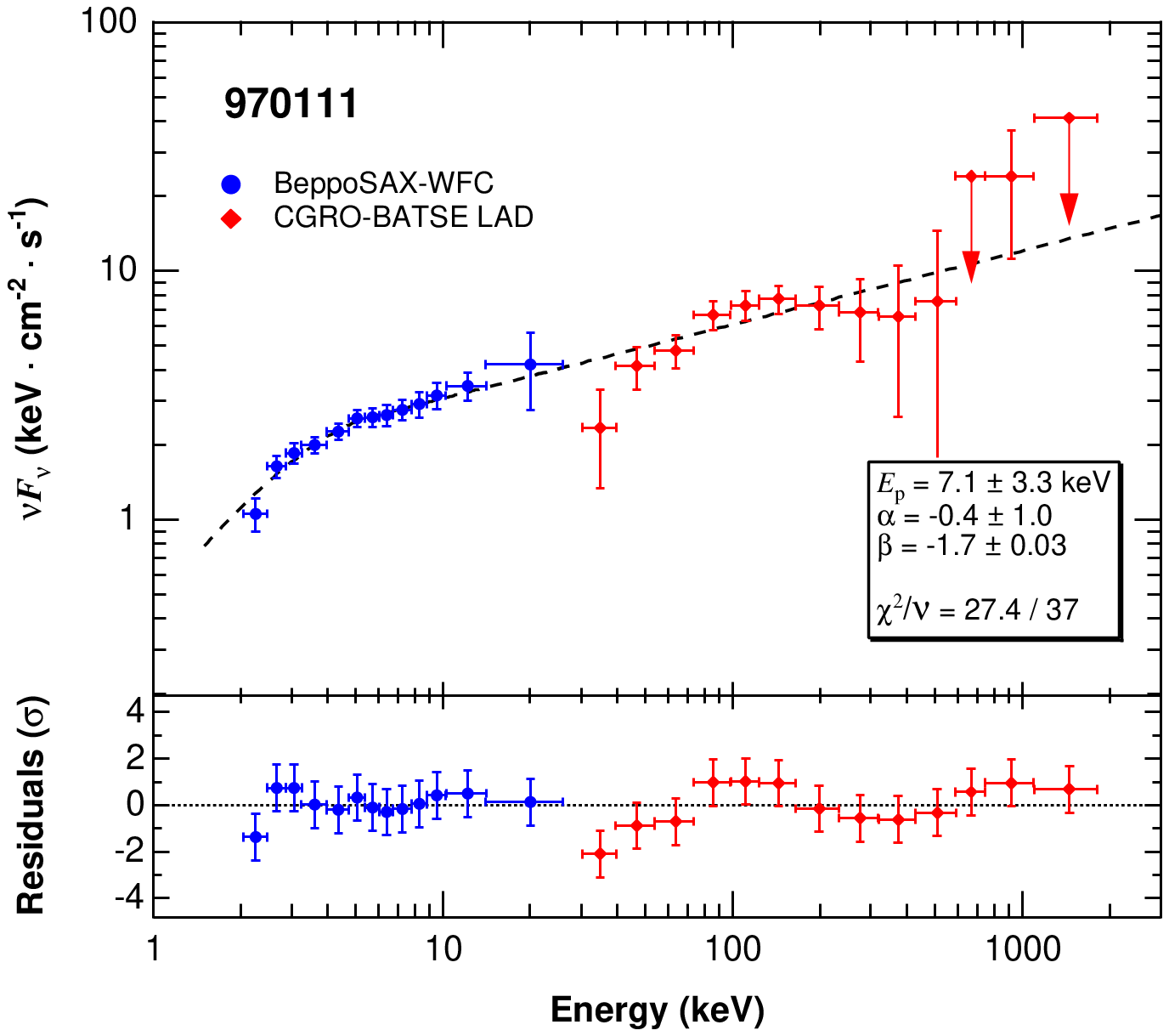}
\caption[]{{\em left:} WFC (upper panel) and BATSE (lower 4 panels)
light curves of the 970111B event. Note that the event was earth
occulted for BATSE prior to -70~s. The dashed lines delimit the time
interval of the spectrum acquired (right) \label{fig970111a}.}
\end{figure}

\section{Individual cases}

\vspace{0.3cm} \noindent {\bf 961229.}  This event lasted 540~s in
X-rays. Data of the Gamma-Ray Burst Monitor on BeppoSAX (Costa et
al. 1998) do not show a simultaneous increase, with an estimated upper
limit to the peak intensity of 5 c~s$^{-1}$. There are no BATSE
data. The 2--28 keV peak flux is 0.2 Crab units and the peak profile
triangular-like.  The interesting feature of this event is that its
3.3 arcmin-radius (99\%-confidence) error region contains the
double-lobed radio galaxy PKS~J0234-0049 at 2.9 arcmin from the WFC
centroid. PKS~J0234-0049 is 89~mJy at 20~cm (Dunlop et al. 1989; White
et al. 1997). There is no redshift determination. We estimate a
1.0$\times10^{-3}$ chance probability for a radio source this bright
or brighter to be in the WFC error box.  We suspect this is a GRB
because it is hard to explain it as an X-ray burst (atypical time
profile), stellar flare (too short) and short-duration accretion event
on a local galactic compact object (the galactic latitude is
-54\deg). Nevertheless, in absence of $\gamma$-ray measurements, we
cannot completely rule it out as an accretion event.

\vspace{0.3cm} \noindent {\bf 970111B.}  This is the brightest event
in our sample.  It occurred 13.9~hr after the famous GRB~970111. In
the WFC data it lasted 2550~s, similar to GRB~971208 although there is
much more structure in the time profile. For part of the 2550~s there
is concurrent BATSE data during which there is a strong detection
lasting 400~s. The 50-300 keV fluence is measured to be
10$^{-5}$~erg~cm$^2$ which puts it in the top 10\% of the BATSE
sample.  The 2--10 keV over 50--300~keV fluence ratio is 0.2 which is
outside the X-ray flash regime (Heise et al. 2001).

\begin{figure}
\plottwo{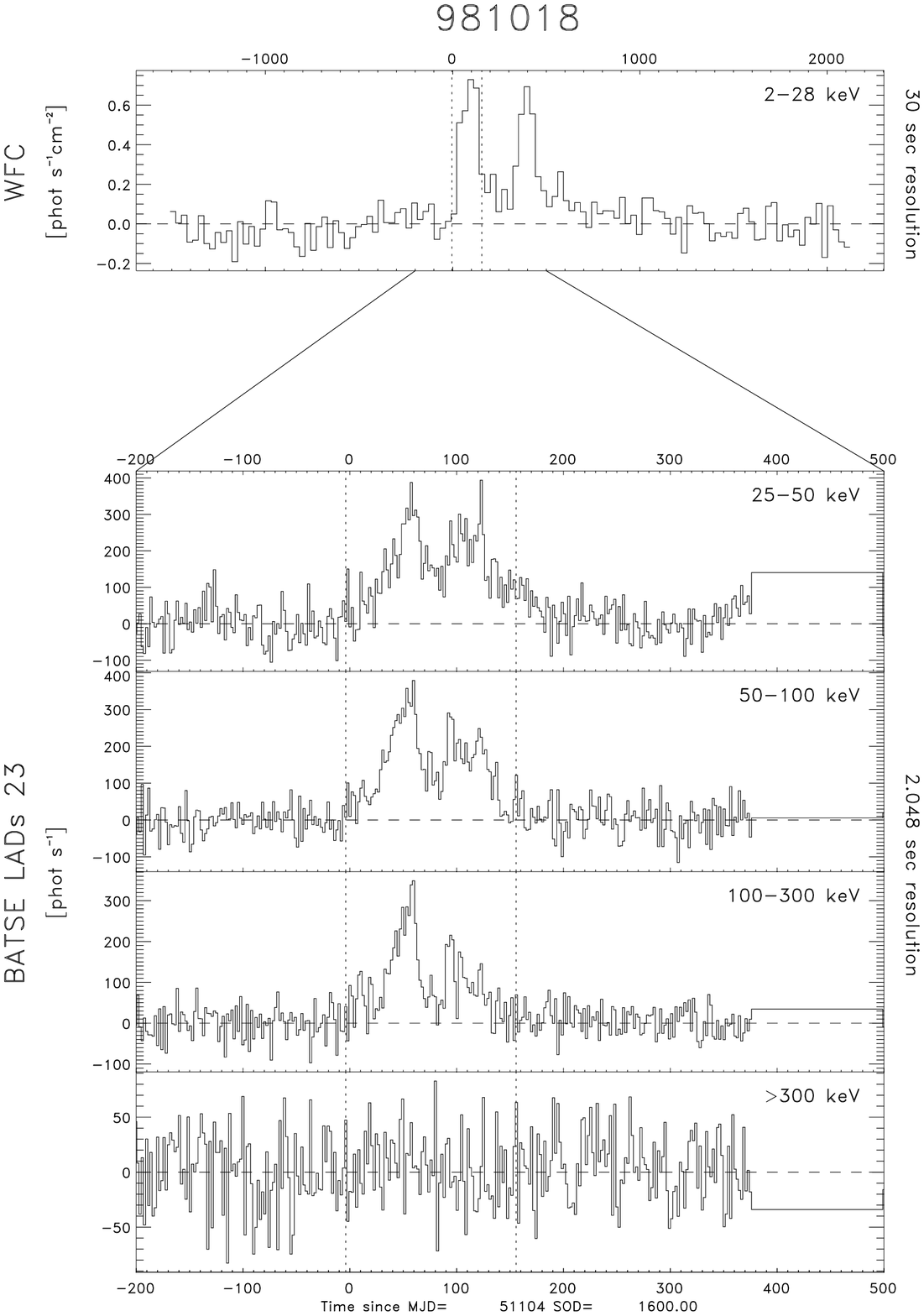}{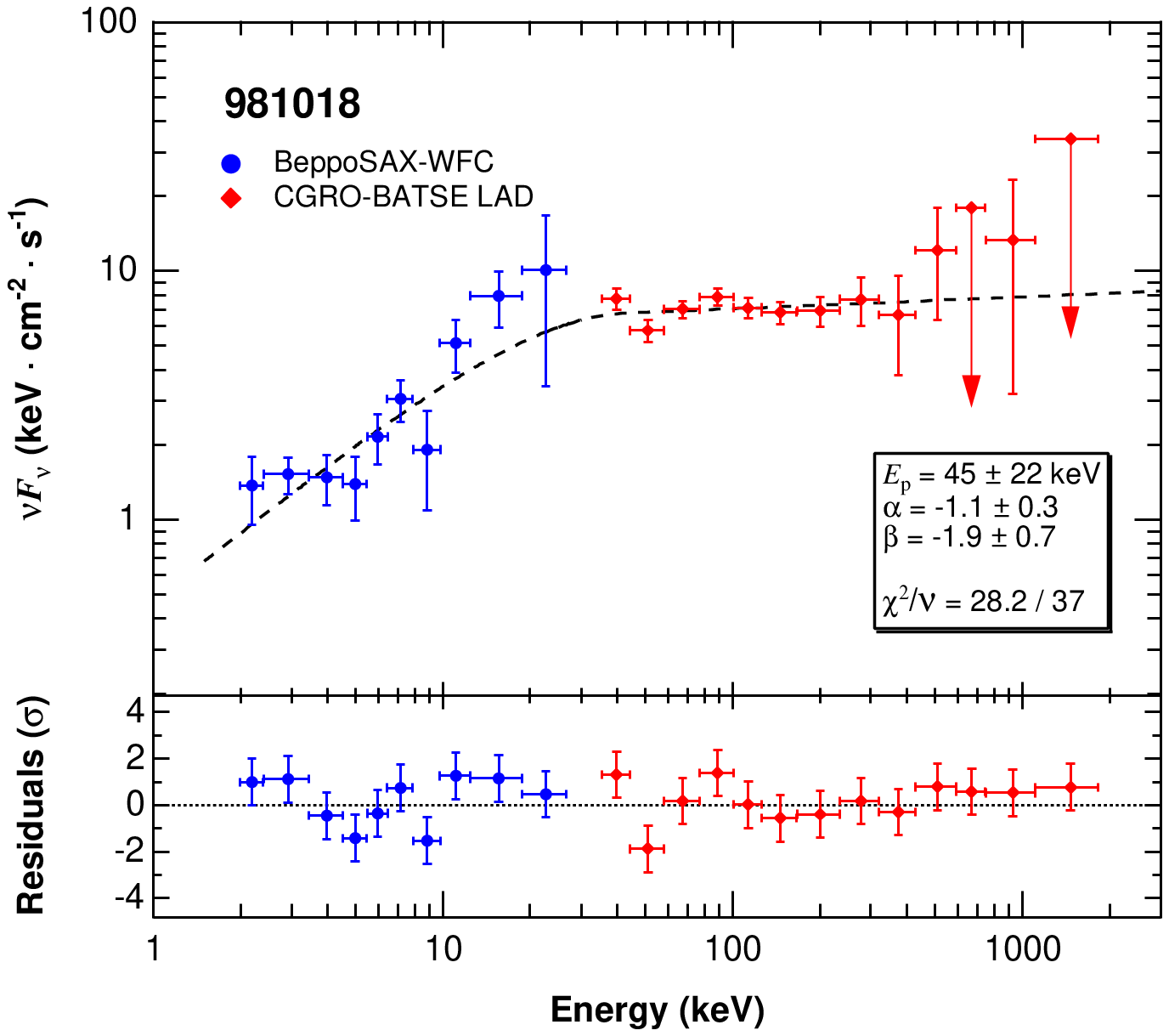}
\caption[]{Light curves and spectrum
of the 981018 event.\label{fig981018}}
\end{figure}

\begin{figure}
\plottwo{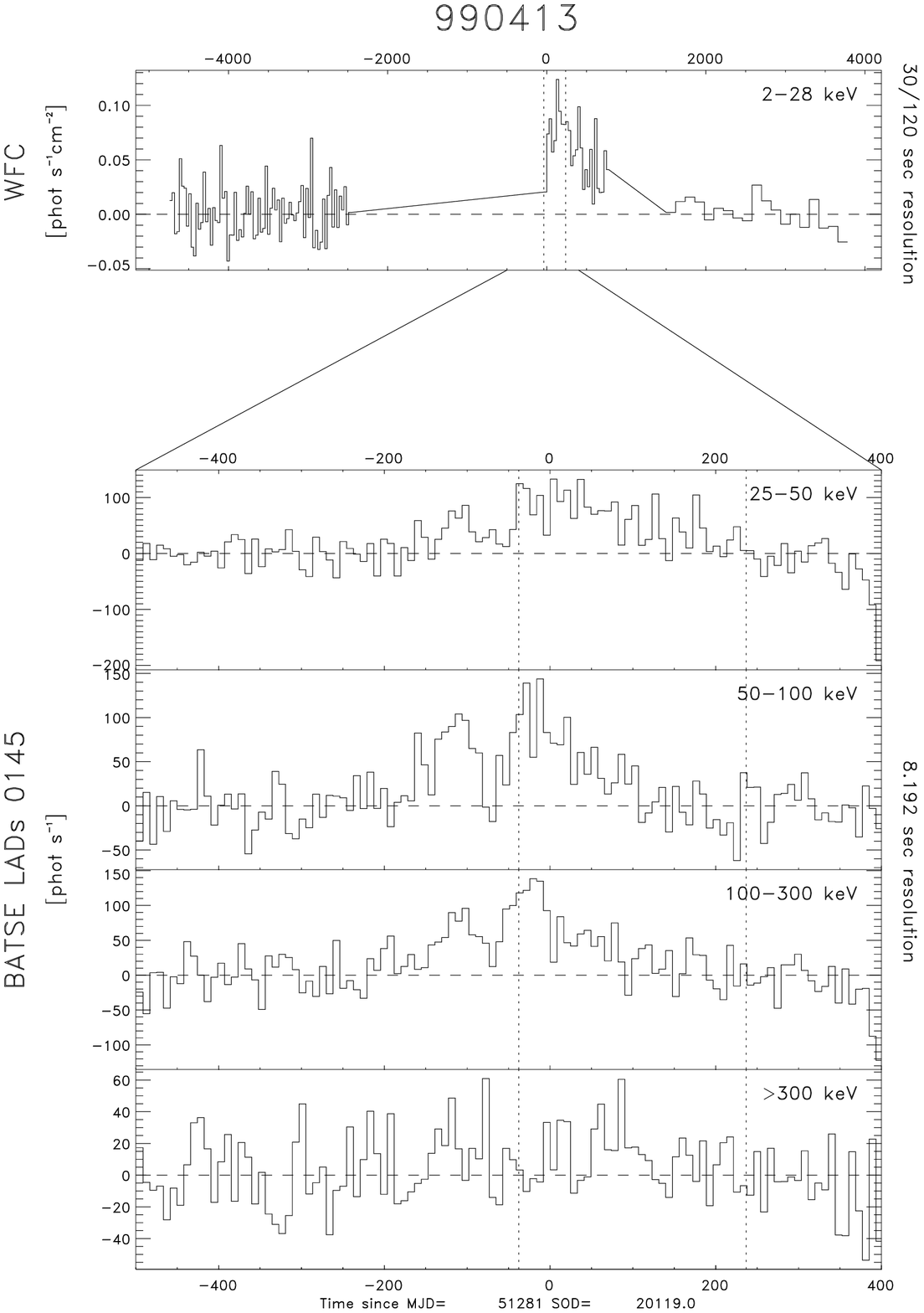}{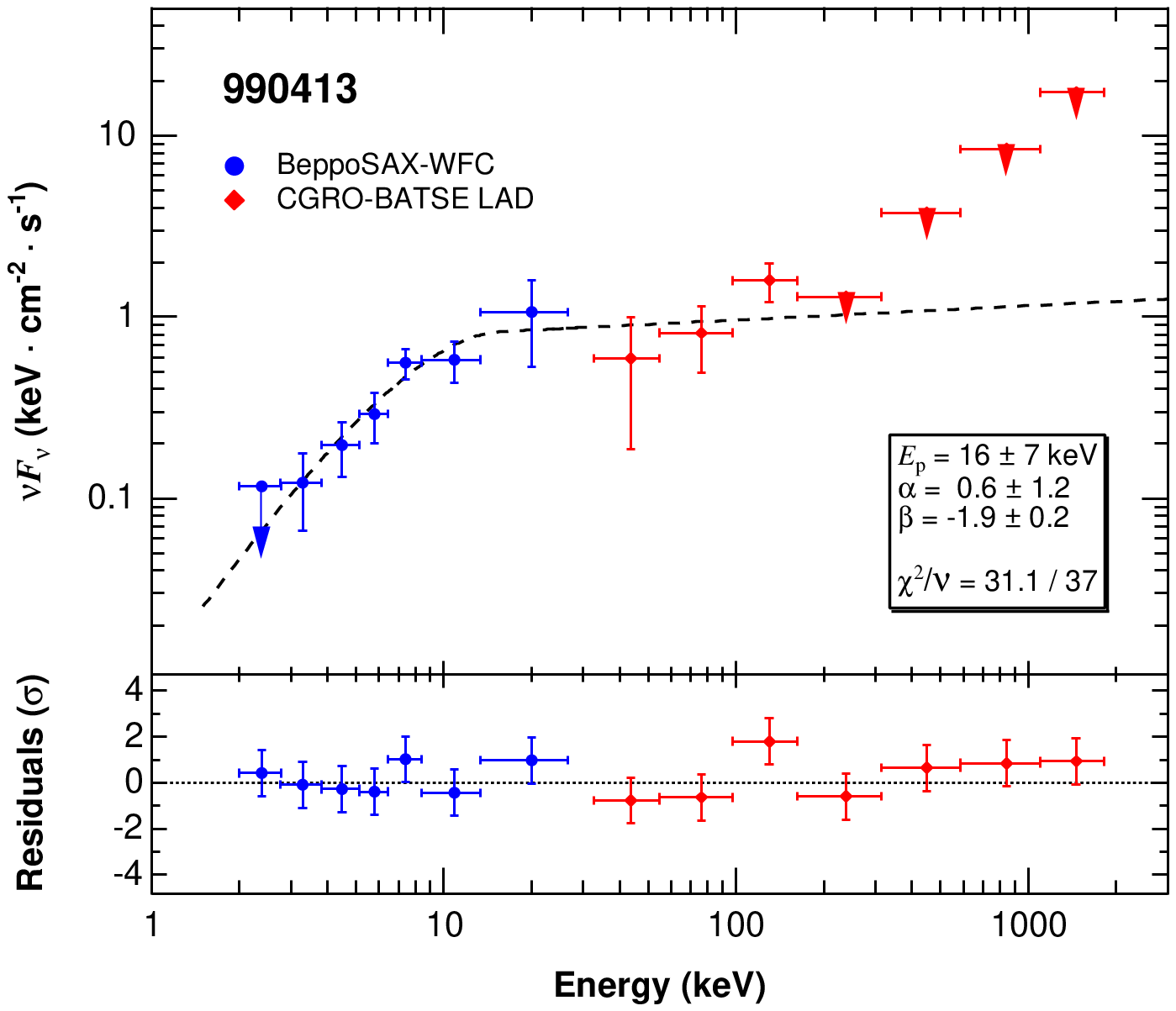}
\caption[]{Light curves and spectrum
of the 990413 event. \label{fig990413}}
\end{figure}

\begin{figure}[t]
\plotfiddle{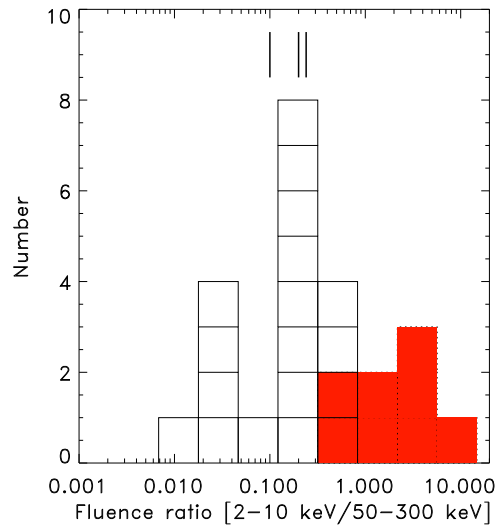}{7.5cm}{0.}{50.0}{50.0}{-150}{-50}
\caption[]{Histogram of X to $\gamma$-ray fluence ratios of 16 GRBs
(open rectangles) and 9 XRFs (grey rectangles) detected with both WFC
and BATSE (Heise et al., in prep.). The ratio for the three long GRBs
discussed in the present paper are indicated by the vertical lines in
the top part of the panel.\label{figfluenceratio}}
\end{figure}

\vspace{0.3cm} \noindent {\bf 981018.}  This again is a fairly strong
detection in both the WFC and BATSE.  The event lasted 900~s and the X
to $\gamma$-ray fluence ratio is 0.24 over 160~s of simultaneous data,
again outside the X-ray flash regime. This is the only event in our
sample which was found independently in searches of untriggered BATSE
events thus far (Stern et al. 2001). It was also detected in an
offline search for GRBs in archival 40-700 keV GRBM data in which it
had a duration of $\approx150$~s (Guidorzi 2002).

\vspace{0.3cm} \noindent {\bf 990413.}  This is a good WFC and BATSE
detection. The X to $\gamma$-ray fluence ratio is 0.10 over 274~s of
simultaneous data. The X-ray duration is ill-determined, the event
being cut on both sides by data gaps; it is between 800 and 4000~s.

\section{Conclusions}

\begin{list}{}{\leftmargin=0.4cm \itemsep=0cm \parsep=0cm \topsep=0cm}
\item[$\bullet$] The longest duration of the prompt X-ray emission from a GRB
(2550~s for GRB~970111B) is similar to that ever measured in
$\gamma$-rays (2000 s for GRB~971208).
\item[$\bullet$] In as far as 2-1000~keV $\nu F_\nu$ spectra could be
measured, they tend to have low peak energies but normal high-energy
power law indices, compared to canonical GRBs. The X-ray to
$\gamma$-ray fluence ratio (see Fig.~\ref{figfluenceratio}) indicates
mildly soft events which, nevertheless, are outside the X-ray flash
regime (cf, Heise et al. 2001).
\item[$\bullet$] Given that two out of three BATSE-detected events
were not identified yet as such, despite the facts that the onboard
trigger algorithm was enabled (on channels 1+2 for 970111B, 3+4 for
981018, 2+3 for 990413) and a number of offline searches have already
been performed (e.g., Stern et al.  2001), there must be many more as
yet unidentified GRBs in BATSE data.
\item[$\bullet$] If the 961229 event, the only one in our sample {\em not}
detected in $\gamma$-rays, is indeed associated to the radio galaxy
PKS~J0234--0049, that would be remarkable and warrants follow-up study
of the galaxy.
\end{list}

\end{document}